\documentclass[12pt]{article}
\usepackage{graphicx}
\usepackage{url}
\usepackage{amssymb}
\usepackage{amsmath}
\usepackage{latexsym}
\newcommand{\lsim}{\!\mathrel{\hbox{\rlap{\lower.55ex \hbox{$\sim$}} \kern-.34em \raise.4ex \hbox{$<$}}}}
\newcommand{\gsim}{\!\mathrel{\hbox{\rlap{\lower.55ex \hbox{$\sim$}} \kern-.34em \raise.4ex \hbox{$>$}}}}
\newcommand{\be}{\begin{equation}}
\newcommand{\ee}{\end{equation}}
\newcommand{\Ogg}{${\mathcal O}_{1}$ }

\begin{document}
\begin{titlepage}
\flushright{HUTP-06/A0037\\
LBNL-61548\\
MCTP-06-20\\}
\vspace{0.3in}
\begin{center}
{\Large \bf Disentangling Dimension Six Operators \\
\vspace{.07in} through Di-Higgs Boson Production}

\vspace{0.5in}
{\bf Aaron Pierce$^{1,2}$, Jesse Thaler$^{3,4}$, Lian-Tao Wang$^{1}$}

\vspace{.5cm}

{\it $^{1}$ Jefferson Laboratory, Physics Department, Harvard University, \\ Cambridge, MA 02138}

\vspace{0.2cm}

{\it $^{2}$ Randall Laboratory, Physics Department, University of Michigan, \\ Ann Arbor, MI 48109}

\vspace{0.2cm}

{\it $^{3}$ Department of Physics, University of California, Berkeley, CA 94720
}

\vspace{0.2cm}

{\it $^{4}$ Theoretical Physics Group, Lawrence Berkeley National Laboratory, Berkeley, CA 94720}
\end{center}
\vspace{0.8cm}

\begin{abstract}
New physics near the TeV scale can generate dimension-six operators that
modify the production rate and branching ratios of the Higgs boson.
Here, we show how Higgs boson pair production can yield complementary
information on dimension-six operators involving the gluon field  
strength.
For example, the invariant mass distribution of the Higgs boson pair can 
show the extent to which the masses of exotic TeV-scale quarks come 
from electroweak symmetry breaking.  We discuss both the current 
Tevatron bounds on these operators and the most promising LHC measurement 
channels for two different Higgs masses: 120 GeV and 180 GeV.  

\end{abstract}

\end{titlepage}
\section{Introduction}

If the Large Hadron Collider (LHC) does not produce any resonances aside from the Higgs boson required to unitarize $W$-$W$ scattering, physicists will be forced to look for new physics in indirect ways.  One approach, recently re-emphasized by \cite{ManoharWise}, is to hunt for new physics via the presence of higher-dimension operators involving only Standard Model fields.  Many of these operators, exhaustively catalogued in \cite{Buchmuller}, are already well constrained by existing precision measurements from LEP and are unlikely to be probed further at the LHC.  Here we discuss higher-dimension operators containing the Higgs boson that are currently poorly constrained, but could directly influence collider phenomenology at the LHC. Our primary focus will be on final states with two Higgs bosons.

Colored particles that get part of their mass from electroweak symmetry breaking (EWSB) can induce the operator
\begin{equation}
{\mathcal O}_{1} = c_{1} \, \frac{\alpha_s}{4 \pi v^2} \, G_{\mu\nu}^a G^{\mu\nu}_a H^{\dagger} H
\end{equation} 
at the loop level.  The mass scale $v = 246$ GeV has been chosen for later convenience, and $4 \pi v$ may or may not be the actual scale of new physics.  The influence of this and other operators on single Higgs boson production and branching ratios was recently discussed in \cite{ManoharWise,ManoharWise2}.  By itself, ${\mathcal O}_{1}$ is insufficient to completely describe the low energy effects on both single and pair Higgs boson production.  To see this, consider a new particle whose mass comes entirely from EWSB.  This yields a different (non-decoupling) operator.  As is familiar from Higgs low energy theorems, a heavy quark with Yukawa coupling $\lambda \rightarrow \infty$ generates not $\mathcal{O}_1$ but
\be
\mathcal{O}_2 = c_2 \, \frac{\alpha_s}{8 \pi} \, G_{\mu\nu}^a G^{\mu\nu}_a \log \left(\frac{H^\dagger H}{v^2}\right),
\ee
which can be understood by thinking of $H$ as a background field and treating the heavy quark mass as a threshold for the running of the QCD gauge coupling \cite{Shifman}.  If we expand $\mathcal{O}_1$ and $\mathcal{O}_2$ in terms of the physical Higgs boson $h$ ($H=\frac{1}{\sqrt{2}}(h + v)$),
\be
\label{eqn:expanded}
\mathcal{O}_1 \supset \frac{c_1 \alpha_{s}}{4\pi} \, G_{\mu\nu} G^{\mu\nu} \left(\frac{h}{v} +  \frac{h^2}{2v^2}\right), \qquad \mathcal{O}_2  \supset \frac{c_{2} \alpha_{s} }{4 \pi} \, G_{\mu\nu} G^{\mu\nu} \left(\frac{h}{v} -  \frac{h^2}{2v^2}\right),
\ee  
then the differing effects on Higgs boson pair production are manifest.

It is also clear that these two operators are sufficient to describe single and pair Higgs production at energies well below the mass scale of the new physics.  General models are effectively described by a linear combination of $\mathcal{O}_1$ and $\mathcal{O}_2$.  A study of Higgs boson pair production would indicate the relative importance of these two operators.  Such an observation would probe the extent to which new colored particles receive their mass from EWSB.



There is even a parametric limit in which deviations arising from the operator $\mathcal{O}_{1}$ might become visible before any new particles are directly observed.  Imagine that there exist $N_f$ new colored particles, all with heavy masses and some coupling to the Higgs boson.  Then the cross section for their direct production scales like $N_f$, but processes that involve the operator $\mathcal{O}_{1}$ will go like $N_f^{2}$.  So, in the limit of large $N_f$, the indirect effects of the operator will be visible first, and the effects described in this letter could be the first indications of new physics. 

Of course, even if the new physics is produced directly, Higgs pair production would remain an interesting channel to help disentangle the new physics.  New physics predictions for Higgs pair production already exist in a model-dependent context, including Little Higgs Models \cite{LHDH}, Randall-Sundrum like models \cite{RSDH}, extended Higgs Sectors \cite{ExtendedHiggs,HigherHOps}, top condensation models \cite{SpiraWells}, and the Minimal Supersymmetric Standard Model \cite{PlehnMSSM,MSSMDH}.  By using the language of $\mathcal{O}_1$ and $\mathcal{O}_2$ we can show the importance of Higgs pair production model-independently.

In the following section, we describe the new physics that can give rise to the operators ${\mathcal O}_{1}$ and ${\mathcal O}_{2}$.  Next, we review the constraints on the possible size of these operators.  We then discuss possible discovery of these operators in Higgs pair production, considering two concrete cases: $m_{h} \sim 120$ GeV and $m_{h} \sim 180$ GeV.  We conclude with a brief discussion of other higher-dimensional operators involving the Higgs boson.  We find that the operators considered here are the ones most likely to give rise to interesting effects at the LHC.

\section{Theoretical Considerations}
\begin{figure}
\begin{center}
\includegraphics[scale=1.0]{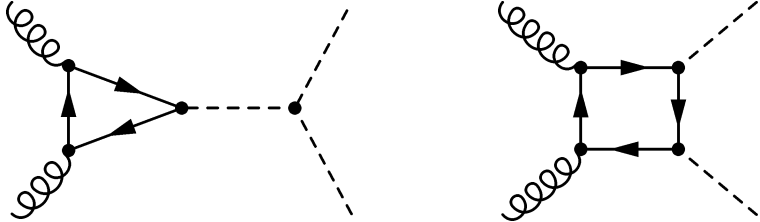}
\end{center}
\caption{The contributions to Standard Model Higgs pair production are dominated by loops containing top quarks.}
\label{Fig:SMDiHiggs}
\end{figure}

The operators \Ogg and ${\mathcal O}_{2}$ have a different dependence on $m_{hh}^2$ than the Standard Model contribution to Higgs pair production.  In particular, amplitudes involving $\mathcal{O}_{1,2}$ will grow like $m_{hh}^2$ all the way up to the mass scale of the new physics.\footnote{In this paper, we will assume that the operators $\mathcal{O}_1$ and $\mathcal{O}_2$ completely dominate the new physics contribution at the LHC, and that higher derivative operators that might soften this behavior are subdominant.} On the other hand, contributions to the amplitude from Standard Model processes shown in Fig.~\ref{Fig:SMDiHiggs} do not grow for energies above the top quark mass.  Assuming the scale of the new physics is much larger than $m_{t}$, this dependence on $m_{hh}^2$ will serve as a way to disentangle the effects of higher-dimension operators from the Standard Model contribution.  

\begin{figure}
\begin{center}
\includegraphics[scale=1.0]{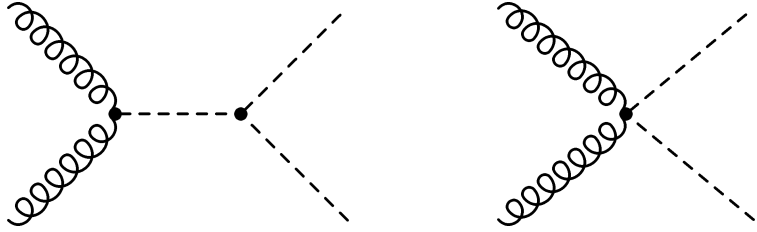}
\end{center}
\caption{The two diagrams that contribute to Higgs pair production coming from the higher dimension operators $\mathcal{O}_1$ and $\mathcal{O}_2$.  In the first diagram, a new $g$--$g$--$h$ vertex combines with the Standard Model three Higgs boson coupling.}
\label{fig:twodiagrams}
\end{figure}

As shown in Eq.~(\ref{eqn:expanded}), \Ogg and ${\mathcal O}_{2}$ induce different contributions to single and pair Higgs boson production. Thus, different kinds of ultraviolet physics can yield different relative signs between the operators containing one and two physical Higgs bosons.  In general, there will be interference between the two diagrams in Fig.~\ref{fig:twodiagrams}.  The amount of interference will give us a handle on the relative size of $c_1$ and $c_2$.  A similar interference is well-known in the Standard Model \cite{Glover} and could potentially be used to measure the $h^3$ Higgs self-coupling \cite{BaurWWPRL, BaurWW}.

There is no renormalizable Lagrangian that can generate $\mathcal{O}_1$ at tree level.   To see how $\mathcal{O}_1$ and $\mathcal{O}_2$ can be generated at loop level, consider the effect of new heavy quarks who get some of their mass from EWSB.  If we treat $H$ as a background field, then each quark mass is a threshold for the one-loop running of the QCD gauge coupling \cite{Shifman}.  Assuming all of the heavy quark masses $m_i \gg m_{h}$, the low-energy QCD gauge coupling is 
\be
\label{eq:gaugerunning}
\frac{1}{g^2(\mu)} = \frac{1}{g^2(\Lambda)} - \frac{b_{UV}}{8 \pi^2} \log \frac{\Lambda}{\mu} - \frac{1}{8\pi^2} \sum_i \delta b_i \log \frac{m_i (\langle H \rangle)}{\mu},  
\ee
where $\delta b_i = 2/3$ for a $SU(3)$ fundamental fermion.  The non-canonically normalized QCD gauge kinetic term
\be
\label{eq:gaugekinetic}
\mathcal{L}_{\rm kinetic} = \frac{-1}{4 g^2 (\mu)} G_{\mu\nu}^a G^{\mu\nu}_a
\ee 
can then be expanded in terms of $h$ to determine the effective values of $c_1$ and $c_2$ relevant for single and pair Higgs production.\footnote{Note that the definitions of ${\mathcal O}_{1}$ and ${\mathcal O}_{2}$ assume canonically normalized kinetic terms.}  At higher order in $h$, ${\mathcal O}_{1}$ and ${\mathcal O}_{2}$ are insufficient to specify all of the allowed Higgs interactions, so Eqs.~(\ref{eq:gaugerunning}) and (\ref{eq:gaugekinetic}) should be used directly.

For concreteness, consider the following Lagrangian ($\tilde{H} = \epsilon \cdot H^\dagger$) 
\begin{eqnarray}
&-\mathcal{L}_{\rm mass} =& \lambda_1 \left(Q H T^c + Q \tilde{H} B^c \right) + \lambda_2 \left(Q^c \tilde{H} T + Q^c H B \right) \nonumber\\
&&~ + m_{A} Q Q^c + m_{B} (T T^c + B B^c) + \mathrm{h.c.},
\end{eqnarray}
where $Q,Q^c$ are vector-like $SU(2)_L$ doublets and $T,T^c$ ($B,B^c$) are vector-like $SU(2)_L$ singlets, with appropriate hypercharges and $SU(3)_C$ couplings.  In order to suppress contributions to the $T$-parameter \cite{PeskinTakeuchi}, we assume custodial isospin.  Using Eq.~(\ref{eq:gaugerunning}) with $i =1$ to $4$ and $\delta b_i = 2/3$:\footnote{In order for this QCD beta function argument to make sense, $\beta$ has to be far from 1, or else there is a mass eigenstate lighter than the physical Higgs.  Note that $(1-\beta)^2$ is proportional to the determinant of the mass matrix.  For simplicity, we take $\beta$ as a real parameter.  In the case where it is complex, the formulae for $c_{1}$ and $c_{2}$ are modified, but they remain real, as Hermiticity of the Lagrangian requires.  When there are phases in the mass matrix, $c_2$ can be negative.}
\be
c_1 = \frac{4}{3} \, \frac{-\beta}{(1-\beta)^2}, \qquad c_2 = \frac{4}{3} \, \frac{1}{(1-\beta)^2}, \qquad \beta \equiv \frac{2 m_A m_B}{\lambda_1 \lambda_2 v^2}.
\ee
If all the mass of the heavy quarks comes from EWSB ($\beta =0$) then $c_1 = 0$.  


Can the effects of $\mathcal{O}_i$ be visible before the new colored states are seen directly?  In the case of heavy quarks that get all of their mass from EWSB, it seems unlikely.  The new quarks could at most have Yukawa coupling $\lambda \sim 4\pi/\sqrt{N_C}$ to keep the theory perturbative, where the number of colors $N_C = 3$ for our toy model.  With masses of $\lambda v/\sqrt{2} \sim 1.3$ TeV, the heavy quarks will have a rather small pair-production cross section $\sim 10$ fb, but could well be visible at the LHC in single production via $b-W$ fusion (see, e.g.\ \cite{HeavyQ}), depending on the flavor structure of the heavy sector.  So, direct production would likely be the first window on new physics of this type.  What about the case where the quarks have mostly vector-like masses ($\beta \gg 1$), so $c_1 \gg c_2$?  The large vector-like mass supresses the overall contribution to $c_{1}$, as this operator decouples like $m^{2}$. So for $c_1$ to be $\mathcal{O}(1)$, the large vector-like mass must be compensated by a large number $N_f$ of heavy quarks: $N_f \sim m^2/\lambda^2 v^2$ to prevent rapid decoupling. Since the $b-W$ fusion process scales roughly as $m^{-7}$, the total production cross section for $N_f$ copies of new physics will scale like $m^{-5}$.  Thus, there is at least a parametric limit where the effect of $\mathcal{O}_1$ is visible before the new heavy quarks are seen directly.   For reasonable values of $\lambda$, $\beta$, and $N_f$, the mass of the new quarks could even exceed the LHC center-of-mass energy while the contribution of these heavy states to $\mathcal{O}_1$ could remain substantial.

\section{Experimental Constraints}

Direct experimental constraints on $\mathcal{O}_1$ and $\mathcal{O}_2$ are quite weak for a low mass Higgs.  Direct constraints come from Tevatron searches for Higgs boson production via gluon-gluon fusion, which constrain the combination $(c_1 + c_2)$.  This production channel is generally not useful at the Tevatron when searching for Standard Model Higgs boson with unmodified properties---the backgrounds are generically too large.  In the low-mass ($m_{h} \lsim 130$ GeV) region, the Higgs boson dominantly decays to $b$ quarks, with the leading sub-dominant decay to $\tau$ leptons.   Decays to $b$ quarks cannot be used for the Higgs boson search because of the too-large QCD background from di-jets.  

A more promising channel is $h \rightarrow \gamma \gamma$.  In the Standard Model, this branching ratio is small, Br$(h \rightarrow \gamma \gamma) \sim 2 \times 10 ^{-3}$.  Current bounds from the DZero experiment constrain the branching ratio to be less than Br($h\rightarrow \gamma \gamma )<$ 0.5 \cite{DZerogammagamma}, assuming the Standard Model production cross section.  To see how this measurement constrains $(c_1 + c_2)$, it is convenient to write the modified cross section times branching ratio for single Higgs production as
\be
\label{eq:prod_decay}
\sigma (i \rightarrow h ) \times {\rm Br}(h \rightarrow f) = \frac{\sigma^{\rm SM} (i \rightarrow h)}{ \Gamma^{\rm SM}_{i}} \frac{\Gamma_{i} \Gamma_f}{\Gamma},
\ee
where $\Gamma_{i,f}$ is the partial width for Higgs decay into the $i$ and $f$ states, and $\Gamma$ is the total width:
\be
\Gamma=\Gamma_{gg}+ \Gamma_{\gamma \gamma}+\Gamma_{WW} + \Gamma_{ZZ} +\Gamma_{bb}+\Gamma_{cc}+ \Gamma_{\tau \tau}+ \cdots .
\ee  
In this language, the DZero constraint can be written as 
\be
\label{eq:dzerobound}
\frac{\Gamma_{gg}}{\Gamma_{gg}^{\rm SM}} \frac{\Gamma_{\gamma\gamma}}{\Gamma} < 0.5.
\ee
If the partial width $\Gamma_{\gamma \gamma}$ is unchanged from the Standard Model, then Eq.~(\ref{eq:dzerobound}) implies no bound on $\Gamma_{gg}$, and therefore no bound on $(c_1 + c_2)$.

In actuality, we expect the new physics that contributes to $\Gamma_{gg}$ to also modify $\Gamma_{\gamma\gamma}$ through operators like
\be
\mathcal{O}_{\gamma\gamma} = c_{\gamma} \, \frac{\alpha_{EM}}{4 \pi v} \, F_{\mu\nu} F^{\mu\nu} h.
\ee
Because $\Gamma_{gg}$ starts to dominate the width when $(c_1 +c_2) < -1.75$ or  $(c_1 +c_2) > 1.05$, for large enough $|c_1 - c_2|$,
the branching fraction to photons will be controlled by the ratio of the color charge to the electric charge of these new states.\footnote{The Standard Model contribution to $\Gamma_{\gamma \gamma}$ from the $W$-loop will be negligible in this regime.}  For particles with top and bottom quark-like quantum numbers, we expect the asymptotic behavior
\be
\label{Eqn:Asymp}
\frac{\Gamma_{\gamma \gamma}}{\Gamma}\sim \frac{\Gamma_{\gamma \gamma}}{\Gamma_{gg}} \rightarrow 10^{-2},
\ee
and Eq.~(\ref{eq:dzerobound}) gives the bound:
\begin{equation}
\label{Eqn:DirectGammaGamma}
-2.8 \lsim (c_{1}+ c_{2}) \lsim 2.1, \qquad  \text{(Direct Search for } gg \rightarrow h \rightarrow \gamma \gamma).
\end{equation}   
To obtain these values, we have taken into account the momentum dependence of the top quark loop (a 6\% effect for a 120 GeV Higgs), and assumed similar NLO K-factors for the Standard Model contribution and the contribution coming from the new operator.  This is likely a good approximation, as the radiative corrections are dominated by interactions involving the initial state gluons.  The asymmetric bounds on $(c_1+c_2)$ show the effect of constructive versus destructive interference with the Standard Model.  




In passing, we note that the search for Higgs bosons in the $h \rightarrow \tau \tau$ channel is unlikely to add any useful new constraints on this operator.  For a $m_{h}\sim 120$ GeV, searches for the $A^{0}$ of the MSSM to a pair of $\tau$ leptons places a limit $\sigma(p \bar{p} \rightarrow h) \times \mathrm{Br}(h \rightarrow \tau^{+} \tau^{-}) \lsim 15$ pb \cite{tautausearches}. The rate for this process in the Standard Model is about $0.056$ pb. The decay rate $h\rightarrow \tau^+ \tau^-$ is dominated by SM tree level process and not sensitive to the class of new physics we consider here. Therefore, we do not expect any constraint on $(c_1 + c_2)$ from Tevatron search of this process. 
A similar story applies to searches for $gg \rightarrow h \rightarrow W W^{*}$ in this mass region: by the time the production cross section is large enough to be observable, the Higgs width is dominated by gluons, and the branching fraction to $W$'s is too small to be observed.  Also, even accounting for possible branching ratio enhancements, the search channel $h \rightarrow \gamma Z$ gives slightly worse bounds than the diphoton channel \cite{Abazov:2006ez}.

When the Higgs is heavier than $160$ GeV, the bounds on $(c_1 + c_2)$ become more severe \cite{TevWW}.  Even accounting for the possible increase of $\Gamma_{gg}$, the dominant decay mode is to a pair of on-shell $W$'s at tree level. Since the total width is dominated by $\Gamma_{WW}$, from Eq.~(\ref{eq:prod_decay}) we see that the rate of this process is proportional to $(c_1 + c_2)$.  The combined bounds from the Tevatron allow an increase in the Higgs production cross section of a factor of 6 over the Standard Model rate, corresponding to 
\begin{equation}
\label{Eqn:DirectWW}
-1.2 \lsim (c_{1}+ c_{2}) \lsim 0.5, \qquad (\mbox{Search for $gg \rightarrow h \rightarrow WW$, $m_{h} \sim 180$ GeV}).  
\end{equation}
Similar bounds apply throughout region $m_{h} > 160$ GeV.  For this size of $c_{1}$, the partial width $\Gamma_{gg}$ does not compete with $\Gamma_{WW}$, so the Higgs branching ratio to $W$ bosons is largely unchanged.

While the above discussion summarizes the state of the direct bounds on $(c_1 + c_2)$, there are potential limits from precision electroweak measurements.  In principle, the presence of $\mathcal{O}_1$ and $\mathcal{O}_2$ need not generate operators at a dangerous level, but it should be noted that heavy states that contribute to $\mathcal{O}_1$ and $\mathcal{O}_2$ could contribute to $S$ and $T$.   Even if we assume custodial isospin, new particles can contribute to the operator 
\begin{equation}
{\mathcal O}_{WB} =\frac{c}{\Lambda^2}  H^\dagger W_{\mu \nu} H B^{\mu \nu},
\end{equation}
which is directly related to the $S$-parameter \cite{PeskinTakeuchi}.  The exact contribution to $\mathcal{O}_{WB}$ will depend on the electroweak quantum numbers of the new particles.  If the new particles have quantum numbers identical to Standard Model quarks, this bound on $S$ will give a fairly strict limit on the potential size of $\mathcal{O}_1$ or $\mathcal{O}_2$.  Currently, electroweak precision measurements constrain $S=-0.13 \pm0.10$ \cite{PDG}.  The contribution of a fourth generation of quarks to ${\mathcal O}_{WB}$ yields
\begin{equation}
\Delta S= \frac{1}{2 \pi}.
\end{equation}
Rescaling the quark masses to force $\Delta S < .10$ implies the following rough bound:  
\begin{equation}
\label{Eqn:Sbound}
\Delta S < .10 \implies c_{i} \lsim 0.4.
\end{equation}
Notice that there is no parametric limit in which we could both have sizable contributions to ${\mathcal{O}}_{1,2}$ and decouple contributions to ${\mathcal O}_{WB}$. For example, in our toy model, $N_f$ will scale like $m^2/\lambda^2 v^2$, which cancels the high scale suppression in front of the $H^\dagger W_{\mu \nu} H B^{\mu \nu}$ operator. 

However, if the new colored states have exotic quantum numbers, there can be a sizable enhancement to the $c_{i}$.  In particular, the contributions to $c_{i}$ goes like the Dynkin index of the $SU(3)$ representation, while the contribution to the $S$-parameter goes like the the dimension of the $SU(3)$ representation.  For exotic representations, this can cause a substantial deviation from the prediction of  Eq.~(\ref{Eqn:Sbound}).\footnote{Sufficiently large representations will cause $SU(3)_{C}$ to become asymptotically non-free, and hit a Landau pole slightly above the mass of the exotic representation.  We remain agnostic as to what new physics lies above the scale where these new operators are generated, so we do not view this constraint as too limiting.}  Furthermore, it is possible that exotic representations of $SU(2)_L \times SU(2)_{R}$ might be present, and these can give contributions to $S$ of either sign \cite{DuganRandall}, so a combination of representations might leave $S$ unchanged while giving a large contribution to $\mathcal{O}_1$ and $\mathcal{O}_2$.  Finally, colored scalars can contribute to $\mathcal{O}_1$ without contributing at all to the $S$ parameter:  the interaction $\phi^\dagger \phi H^\dagger H$ does not require $\phi$ to have any electroweak quantum numbers.  So, while $c_{i} \sim 0.4$ may represent a typical value in some theories, values much larger are certainly possible, so we will consider the $c_{i}$ as free parameters up to the limits imposed by the direct searches, Eqs.~(\ref{Eqn:DirectGammaGamma}) and (\ref{Eqn:DirectWW}).  


\section{Measuring Higgs Pair Production}

The sizes of $c_1$ and $c_2$ could be determined uniquely by measuring the cross sections $\sigma(pp \rightarrow h)$ and $\sigma(pp \rightarrow h h)$.  However, the quantity that is most directly measured experimentally is not the cross-section, but rather a cross-section times a branching ratio.  Given the likely possibility that new physics will modify the Higgs branching ratios, it is useful to have an independent measure of $c_1$ and $c_{2}$.

A differential distribution of the form
\be
\frac{d \sigma}{d x } = f(c_1,c_2) \times \frac{\Gamma_{f_1}}{\Gamma} \times
\frac{\Gamma_{f_2}}{\Gamma}, 
\ee 
where $x$ is some kinematical variable, such as $m_{hh}^2$, gives such a handle. In general, $f(c_1,c_2)$ is a function that depends on the size of $c_1$,
$c_2$, their relative sign, as well as interference with the Standard Model piece.  Therefore, both the rate and the shape of the di-Higgs distribution give independent probes of the coefficients $c_1$ and $c_2$, with different systematic errors associated with each measurement.



Whether the shape of the di-Higgs invariant mass distribution can be observed or not depends on the  overall rate of Higgs pair production at a given luminosity as well as the Standard Model background to the channel used to reconstruct the Higgses.  In the following subsections, we comment on two different Higgs mass windows:  a low-mass Higgs boson near 120 GeV and a Higgs boson above the $W^+W^-$ threshold near 180 GeV.  To investigate this question, we augmented MadGraph \cite{MadGraph} with new HELAS \cite{HELAS} routines to simulate the contributions of both the Standard Model top quark loop \cite{Glover,Dicus,PlehnMSSM} and the operators $\mathcal{O}_1$ and $\mathcal{O}_2$.\footnote{We thank Rikkert Frederix for providing the HELAS routines to implement the $G_{\mu\nu} G^{\mu\nu} h$ vertex.}  The FF package \cite{FF} was used in the numerical evaluation of the relevant top loop integrals.  In the limit where the contributions of our new operators vanish, our numerical results agree with those of \cite{PlehnMSSM}.

A search for $\mathcal{O}_1$ and $\mathcal{O}_2$ would involve the same final states as the search for the $h^3$ Higgs self coupling \cite{BaurWWPRL, BaurWW}, but typical values of $c_1$ and $c_2$ can lead to an order of magnitude increase in the cross section $\sigma(gg \rightarrow hh)$  (and more extreme values of $c_i$ can give enhancements of even larger factors that are nevertheless consistent with all known phenomenological constraints). A preliminary ATLAS \cite{ATLASstudy} study suggests that the Higgs self coupling could only be measured with luminosities typically associated with the SLHC.   It would be interesting to know whether the possible drastic increase in cross section from dimension six operators would paint a more optimistic picture for Higgs pair production at the LHC even after accounting for detector resolution effects.  

\subsection{Low Mass Region}
In the region where the Higgs has a low mass $m_{h} \sim 120$ GeV, two Higgs decays can best be observed via the process $gg \rightarrow h h  \rightarrow b \bar{b} \gamma \gamma$ where at least one of the jets is $b$-tagged \cite{PlehnRare}.  While this process has manageable background, in the Standard Model it suffers from the small branching ratio Br$(h\rightarrow \gamma \gamma) \sim 2 \times 10^{-3}$.  As discussed in the previous section, both the branching ratio to photons and overall production rate can be substantially changed from the Standard Model prediction.  However, the branching ratio to $b$-quarks will generically decrease because of the increased partial width $\Gamma_{gg}$, and at the extremes of the Tevatron allowed region in Eq.~(\ref{Eqn:DirectGammaGamma}), Br($h \rightarrow b \bar{b}$) decreases by about factor of four.  Together with a possible factor of five increase in Br($h \rightarrow \gamma \gamma$) described in Eq.~(\ref{Eqn:Asymp}), the overall branching ratio of $hh \rightarrow b \bar{b} \gamma \gamma$ could be comparable to the Standard Model value.  Away from the extreme values of $c_{i}$, it is possible that the product Br($h \rightarrow gg$)$\times$ Br($h \rightarrow \gamma \gamma$) could be larger or smaller than the Standard Model, depending on the extent of cancellation between the new physics and the $W$-loop contribution to the $h \rightarrow \gamma \gamma$ rate.

\begin{figure}
$$
\begin{tabular}{r|rrrrrrrrr}
& $c_1 = -3.0$ & $-2.0$& $-1.0$ & $-0.5$ & $0.0$ & $0.5$ & $1.0$ & $2.0$ & $3.0$ \\
 \hline
$c_2 = -3.0$& 48& 16& 33& 60& 95& 150& 210& 380& 590\\
$-2.0$& 80& 21& 8.6& 21& 46& 82& 130& 270& 450\\
$-1.0$& 130& 44& 4.4& 2.2& 13& 35& 71& 180& 330\\
$-0.5$& 170& 48& 9.8& 1.2& 4.7& 20& 49& 140& 280\\
  $0.0$& 200& 87& 18& 4.2& 1.0& 8.6& 30& 110& 210\\
$0.5$& 230& 120& 36&13& 2.3& 3.9& 8.9& 82& 190\\
$1.0$& 300& 150& 56& 26& 8.6& 3.2& 9.9& 60& 160\\
$2.0$& 410& 240& 110& 68& 37& 16& 6.2& 31& 100\\
$3.0$& 540& 340& 190& 130& 84& 51& 27& 6.2& 64
\end{tabular}
$$
\caption{The ratio of $\sigma(gg \rightarrow hh)$ to the Standard Model di-Higgs cross section for $m_h = 120$ GeV.  This includes the effect of interference between the contributions from $\mathcal{O}_1$, $\mathcal{O}_2$, and the Standard Model.  We assume the new contribution to di-Higgs production inherit the same NLO K-factors as the Standard Model.  The Standard Model cross section is 30 fb, and the allowed range in Eq.~(\ref{Eqn:DirectGammaGamma}) from direct Tevatron searches is $-2.8 \lsim (c_1 + c_2) \lsim 2.1$.}
\label{fig:120table}
\end{figure}

The enhancement of $\sigma(gg \rightarrow hh)$ relative to the Standard Model value for a range of values $c_1$ and $c_2$ is given in Fig.~\ref{fig:120table}.  Each tree-level cross section is multiplied by a K-factor of 1.65 to take into account NLO effects \cite{NLOCorrection}, where we are assuming that the QCD corrections to the diagrams in Figs.~\ref{Fig:SMDiHiggs} and \ref{fig:twodiagrams} are comparable.  The factorization scale and the renormalization scale are taken to be $m_{hh}$, and the CTEQ6L1 parton distribution functions \cite{CTEQ} are used.  As has been previously noted, the dependence on the renormalization scale is somewhat severe for this case, and can easily affect the cross section by tens of percent.

In \cite{PlehnRare}, it was found that with a set of reasonable cuts to isolate the signal, there could be 6 signal events compared to 14 background events with 600 fb$^{-1}$ of LHC data.  Given the large possible enhancements of the rate of Higgs pair production in Fig.~\ref{fig:120table}, the signal to background ratio can be much enhanced, and the event rate need not be so tiny.  This assumes that $\mathcal{O}_{1}$ and $\mathcal{O}_{2}$ do not make large contributions to the background.  In particular, one might worry that an enhanced $g$--$g$--$h$ vertex could influence the irreducible background from $hb\bar{b}$ and $h\gamma \gamma$.  However, as shown in $\cite{PlehnRare}$, these backgrounds are negligible compared to the background from QCD with fakes, so it is reasonable to expect that the values in Fig.~\ref{fig:120table} represent the real increase in the signal to background ratio except at the most extreme values of the $c_i$.


\begin{figure}
\begin{center}
\includegraphics[width=4in]{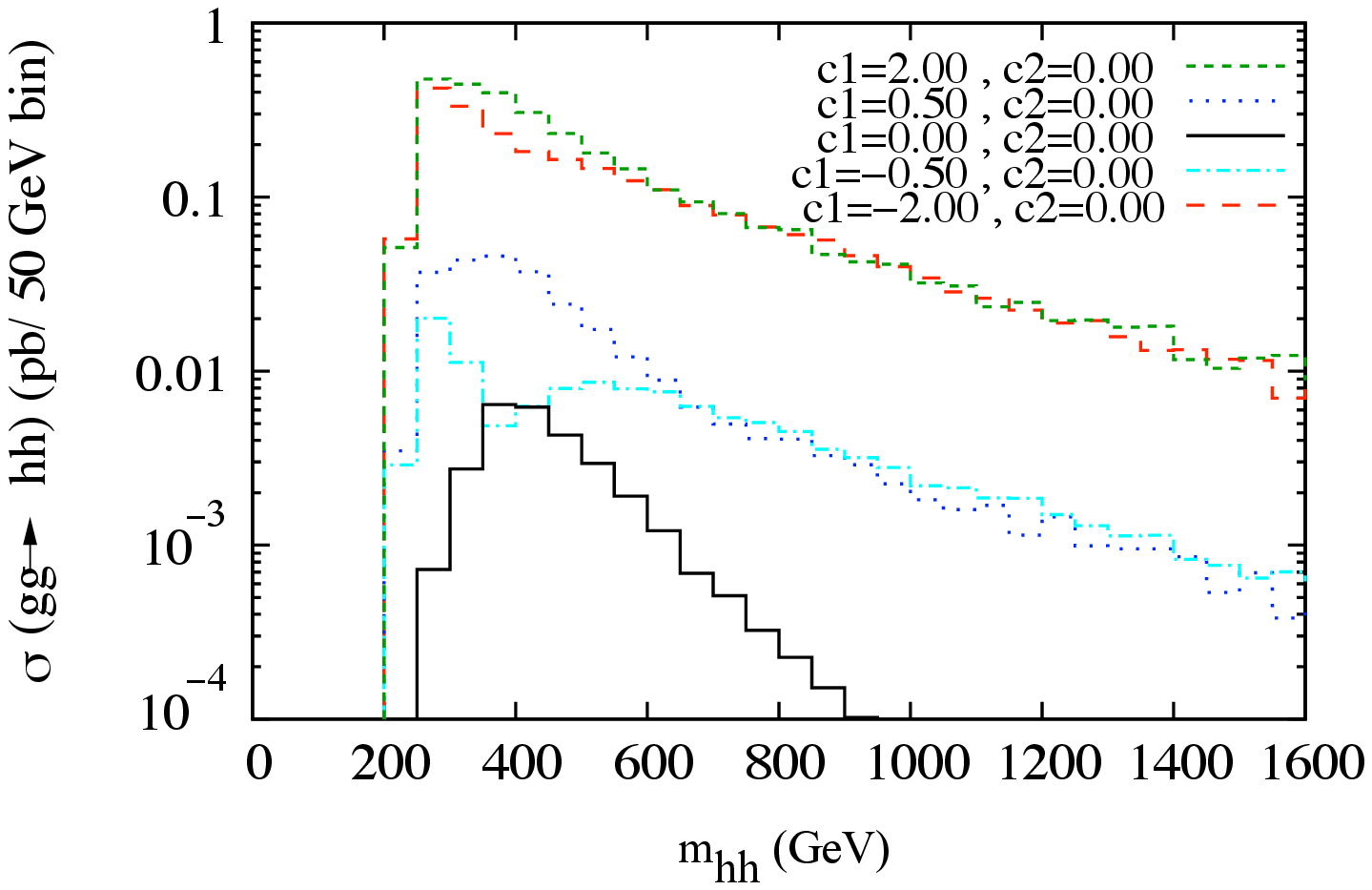}

~\\

\includegraphics[width=4in]{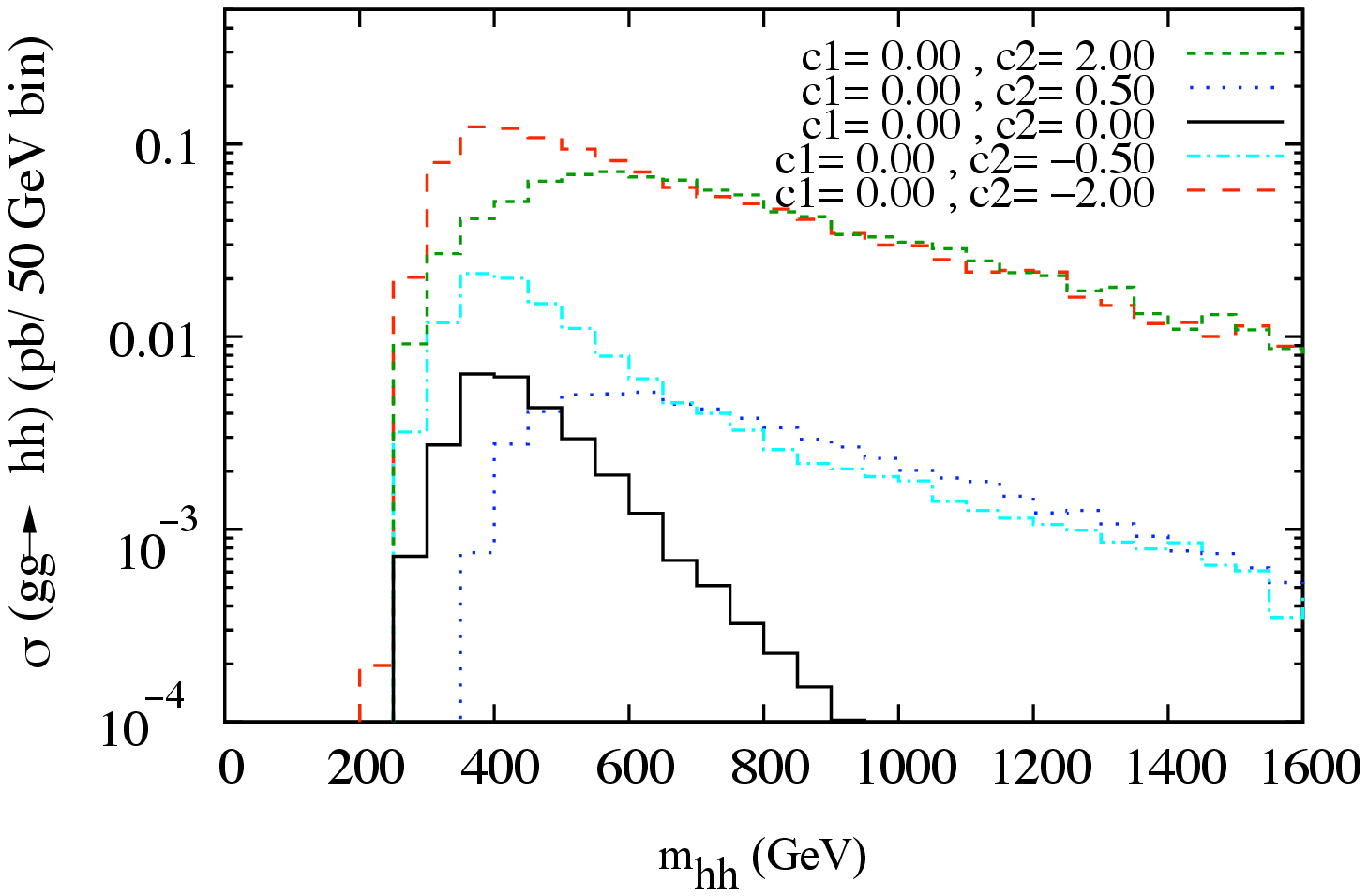}

\end{center}
\caption{Differential cross-sections as a function of $m_{hh}$.  In the top graph $c_2 = 0$ is fixed while $c_{1}$ varies, and in the bottom graph $c_1 = 0$ is fixed while $c_{2}$ varies.  We have set $m_{h}= 120$ GeV and $m_{t}=174.3$ GeV.   The curves for $m_h = 180$ GeV are quite similar, with the trivial modification that the threshold energy is changed.   Note that the asymptotic behavior at large $m_{hh}$ is controlled by the difference $|c_1 - c_2|$.  When $c_1 = -0.5$ and $c_2 = 0$, there is a pronounced dip at $m_{hh} = 400$ GeV, coming from interference between $\mathcal{O}_1$ and the Standard Model top loops.}
\label{fig:diffxsec1}
\end{figure}

\begin{figure}
\begin{center}
\includegraphics[width=4in]{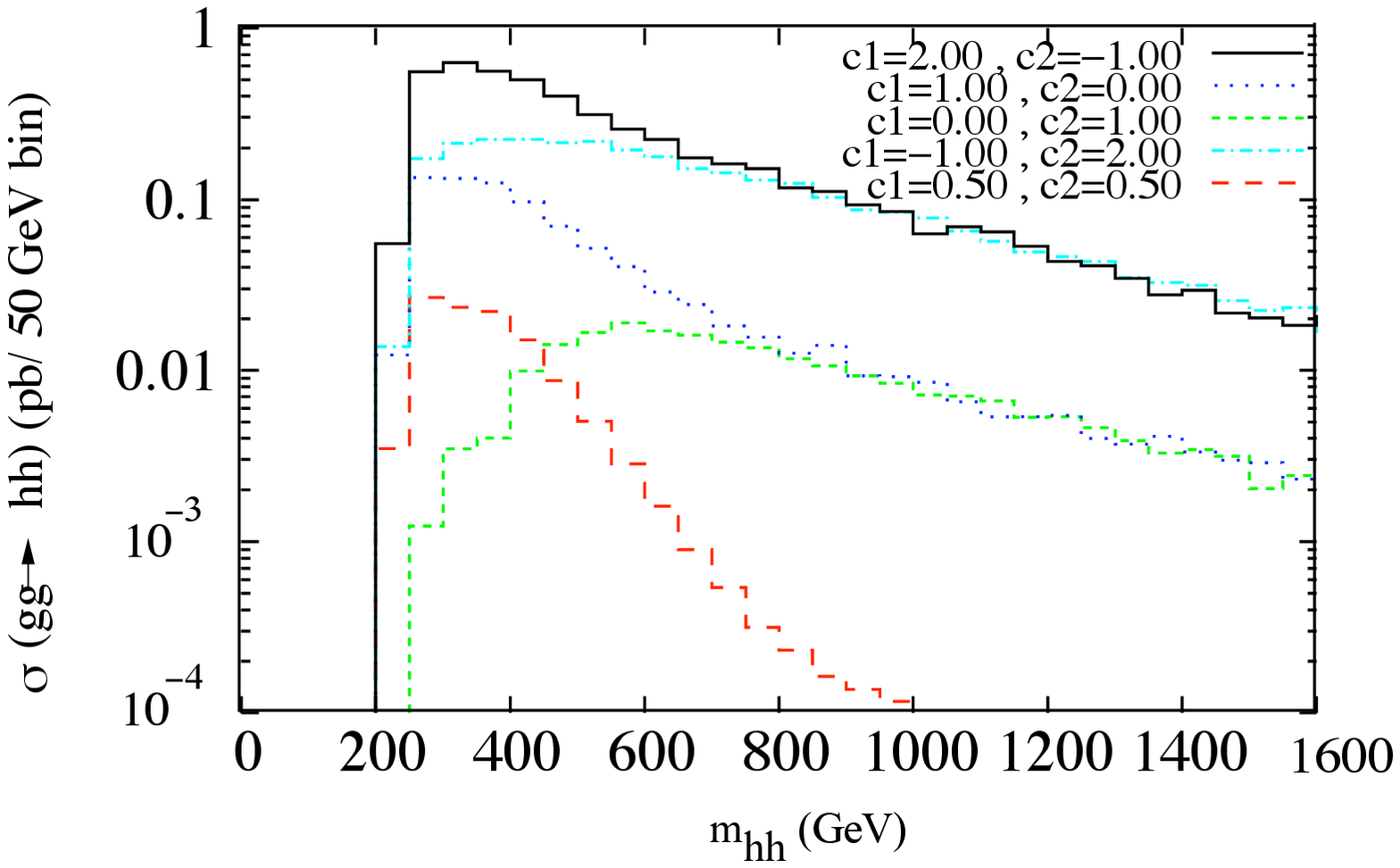}
\end{center}
\caption{Differential cross-sections as a function of $m_{hh}$ for $m_{h}= 120$ GeV.  Here, the single Higgs production rate is fixed by fixing $(c_1 + c_2)$, but the properties of di-Higgs production are clearly modified as the proportion of $\mathcal{O}_1$ and $\mathcal{O}_2$ changes.}
\label{fig:diffxsec2}
\end{figure}

The differential cross sections as a function of $m_{hh}$ are shown for various choices of the $c_{i}$ in Figs.~\ref{fig:diffxsec1} and \ref{fig:diffxsec2}.  Because the rate of single Higgs production is related to $(c_1 + c_2)$, we keep the sum $(c_1 + c_2)$ fixed in Fig. \ref{fig:diffxsec2} while varying the difference.  The large variety of shapes and normalizations show how di-Higgs production can be an important probe of the properties of the Higgs boson.  There are two specific features to notice in these figures.  First, compared to the Standard Model alone, there is a long tail in the $m_{hh}$ distribution when $\mathcal{O}_i$ is turned on, and the size of that tail is governed by $|c_1 - c_2|$.  Second, for values of $c_1$ and $c_2$ that are not too large ($\lsim$ 0.5), interference with the Standard Model can be important, possibly causing a significant deficit of events around $m_{hh} = 400$ GeV.

With sufficient luminosity, a measurement of $\sigma(gg\rightarrow hh)$
as well as of the $m_{hh}$ distribution in di-Higgs events will give some handle on the sizes of $\mathcal{O}_1$ and $\mathcal{O}_2$.  It is beyond the scope of this paper to try to estimate the errors on the measurements of $c_1$ and $c_2$ due to energy resolution, background subtraction, and statistics.  However, because of the large enhancement of di-Higgs production due to these new operators, it is not unreasonable to expect some $m_{hh}$ shape information to be available after several years of high luminosity (100 fb$^{-1}$/yr) running at the LHC.
 
\subsection{Above the $W^{+}W^{-}$ Threshold}
For a Higgs with a mass near 180 GeV, its dominant decay mode is to $W^+W^-$.  The $h h \rightarrow W^+ W^- W^+ W^-$ mode was considered in \cite{BaurWWPRL,BaurWW}, where it was determined that the the cleanest channel for discovering Higgs pairs was when the four $W$'s yielded two same sign leptons, i.e.  $h h \rightarrow W^+W^-W^+W^- \rightarrow jjjj \ell^\pm \nu \ell^\pm \nu$.  For various values of $c_i$, the cross sections in this channel are given in Fig.~\ref{fig:180table}.  Using the cuts in \cite{BaurWW} with 600 fb$^{-1}$, there are 110 Standard Model di-Higgs events with an expected signal/background ratio of 1 to 5.

\begin{figure}
$$
\begin{tabular}{r|rrrrrrr}
& $c_1 = -2.0$& $-1.0$ & $-0.5$ & $0.0$ & $0.5$ & $1.0$ & $2.0$\\
 \hline
$c_2 = -2.0$& 44& 10& 21& 55& 100& 180& 380\\
$-1.0$& 86& 11& 2.7& 14& 46& 97& 260\\
$-0.5$& 120& 21& 2.9& 4.3& 25& 66& 200\\
$0.0$& 150& 39& 9.9& 1.0& 12& 36& 170\\
$0.5$& 200& 63& 24& 4.5& 5.0& 26& 130\\
$1.0$& 260& 94& 44& 15& 5.0& 15& 95\\
$2.0$& 370& 170& 110& 54& 25& 15& 53
 \end{tabular}
$$
\caption{The ratio of $\sigma(gg \rightarrow hh)$ to the Standard Model di-Higgs cross section for $m_h = 180$ GeV. The Standard Model cross section is 13 fb, and the bounds in Eq.~(\ref{Eqn:DirectWW}) from direct Tevatron searches is $-1.2 \lsim (c_{1}+ c_{2}) \lsim 0.5$.}
\label{fig:180table}
\end{figure}

Because of the presence of two final-state neutrinos, the invariant mass of the two Higgses cannot be determined uniquely.  In \cite{BaurWW}, an approximate variable $m_{\rm visible}$ was used, which systematically underestimates the real invariant mass by taking the invariant mass of just the final state leptons and jets.  On the other hand, once jets are combined to give pseudo-$W$'s, the decay topology of $h h \rightarrow W^+W^+ \ell^- \nu \ell^- \nu$ allows for a full reconstruction of the neutrino four-vectors, up to discrete ambiguities. Eight constraints are required to measure the eight components of the two neutrino four-vectors.  They are obtained by requiring that the two neutrinos are on-shell (2), that the neutrinos and charged leptons must reconstruct two $W$'s (2), that opposite sign $W$ pairs must reconstruct two Higgses (2), and that the transverse momenta of the neutrinos must yield the missing $p_T$ vector (2).

\begin{figure}
\begin{center}
\includegraphics[width=3.6in]{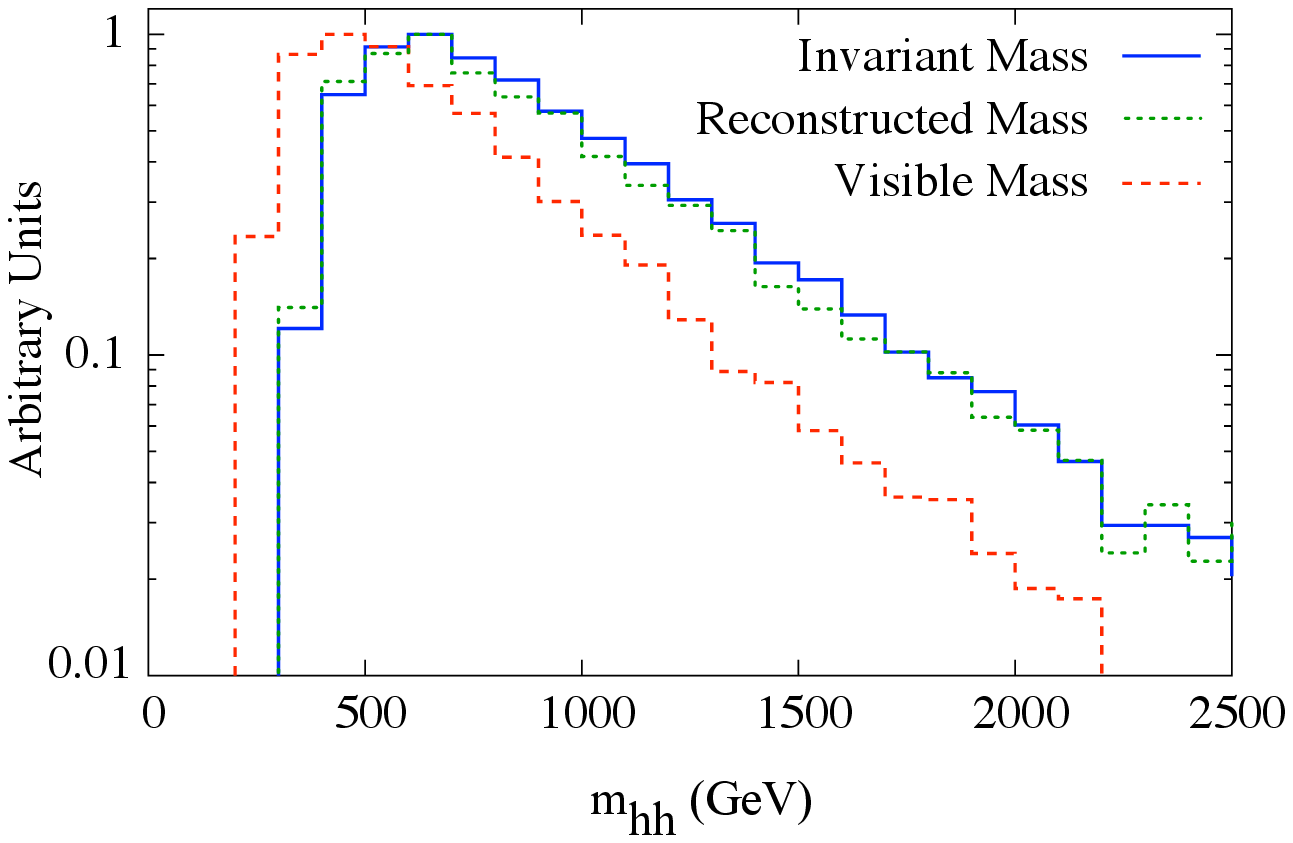}

~\\

\includegraphics[width=3.6in]{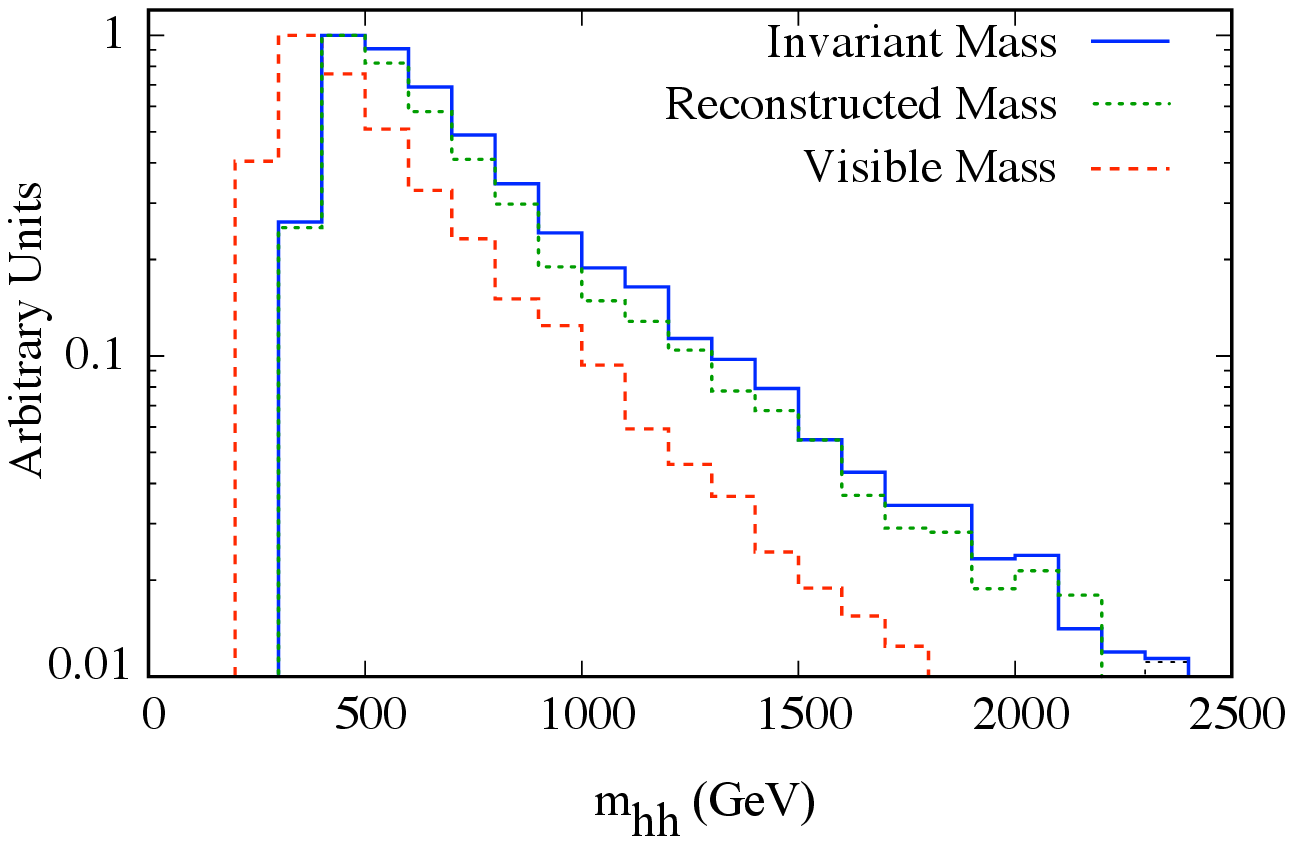}
\end{center}
\caption{Distributions of $m_{hh}$ at $m_h = 180$ GeV for (top) $c_1=0$, $c_2=0.5$ and (bottom) $c_1 = -0.5$, $c_2 = 0$.  The neutrino reconstruction technique does a better job at matching the real invariant mass distribution than the $m_{\rm visible}$ variable, at the expense of reducing the statistics by a factor of two because of the 20\% failure rate for finding any solution to the constraint equations as well as our \emph{ad hoc} method for resolving reconstruction ambiguities.  In these plots, the curves are normalized to peak at 1 in arbitrary units.  Note that regardless of whether one uses the visible or reconstructed mass, the shape difference between the top and bottom curves is still observable, so the main reason to use the reconstruction technique is to potentially reduce backgrounds.}
\label{fig:recomass}
\end{figure}

Of course, finite width effects, initial and final state radiation, and energy resolution issues will blur the di-Higgs invariant mass distribution.  Even with a perfect understanding of those experimental issues, there can be anywhere from zero to eight real solutions to the constraint equations,\footnote{Zero solutions arise if enough of the Higgses and $W$'s are sufficiently off-shell.  In this case, the kinematics of the event will not be consistent with, say, the central value of the $W$ mass of 80.3 GeV.  To get eight solutions, note that both the $W$ and Higgs reconstruction equations are quadratic, yielding a maximum of four real solutions, and there is a two-fold ambiguity as to which hadronic $W$ should be paired with which leptonic $W$.} and one must pick a method to choose the right neutrino four-vectors.  Here, we consider an \emph{ad hoc} method to determine $m_{hh}$ that seems to give reasonable results with high efficiency:  each solution to the constraint equations yields a different value of $m_{hh}$, and we plot the mean value of $m_{hh}$ in events where the standard deviation of $m_{hh}$ is less than 10\% of the mean.  This technique removes around half of the candidate events, and ensures that only values close to the true values are plotted.   As shown in Fig.~\ref{fig:recomass}, this reconstruction method does not appear to systematically distort the $m_{hh}$ distribution and gives a much better estimate of invariant mass distribution compared to the $m_{\rm visible}$ variable.

One of the important benefits of this reconstruction technique could be on background reduction.  One of the largest sources of background comes from $pp \rightarrow W h j j$ where the two jets appear to reconstruct a $W$.  In the signal sample, the failure rate for the neutrino reconstruction (\emph{i.e.}\ when there are zero solutions to the constraint equations) is around 20\%.  However, in a background sample of $pp \rightarrow W h j j$ where the invariant mass of the two jets (ignoring smearing) are forced to lie within 10 GeV of the $W$ mass, the failure rate is nearly 80\%.  This is likely because the largest contribution to $pp \rightarrow W h j j$ comes from diagrams involving Higgs-strahlung off of $W$'s, which has a very different kinematic structure from true Higgs pair production.  It would be interesting to see whether this four-fold improvement of the signal to background ratio persists in a more realistic background study. 

As mentioned in Figs.~\ref{fig:diffxsec1} and \ref{fig:diffxsec2}, the theoretical  $m_{hh}$ distribution at $m_h = 180$ GeV is qualitatively similar to the $m_h = 120$ GeV case apart from the different kinematic threshold.  It is beyond the scope of this paper to consider the full effect of energy resolution and QCD radiation on Higgs pair reconstruction.  Because of the large number of final states and the need to accurately know the missing $p_T$ vector, neutrino reconstruction may or may not be the best experimental technique for extracting information about the di-Higgs $m_{hh}$ distribution.  Most likely, the most appropriate technique will depend on the total number of events seen.

\section{Conclusions}
If new colored resonances are observed at the LHC, the signature explored in this paper will be useful to explore the extent to which these new particles get their mass from EWSB.  If we are unlucky at the LHC and new particles are inaccessible directly but just around the corner, then the di-Higgs invariant mass distribution can be used to probe them indirectly, possibly providing the first measurement of physics beyond the Standard Model.

One might wonder whether there are other operators that could contribute to Higgs pair production at the LHC.  For example, consider the following dimension six operators from \cite{Buchmuller}: 
\be
{\mathcal O}_{FN} = \psi H \chi^c H^\dagger H,  \qquad {\mathcal O}_{Z'} = \bar{\psi} \sigma_\mu \psi H^\dagger D_\mu H,
\ee
where $\psi$ represents a Standard Model $SU(2)_{L}$ doublet, and $\chi^{c}$ a singlet. We might expect to generate ${\mathcal O}_{FN}$ in some Froggatt-Nielsen \cite{Frog} model, and ${\mathcal O}_{Z'}$ could arise from integrating out a $Z'$ gauge boson that coupled to Standard Model fermions and the Higgs.  However, both of these operators are constrained by existing new physics searches.  Unless we assume minimum flavor violation, then ${\mathcal O}_{FN}$ can introduce dangerous flavor-changing neutral currents, but minimal flavor violation implies that the coefficient of this operator will be tiny for the first two generations.  Similarly, setting the Higgs to its vacuum expectation value in ${\mathcal O}_{Z'}$ will generate anomalous couplings between Standard Model fermions and the $Z$, which are well-constrained by LEP.    

At best, we could consider ${\mathcal O}_{FN}$ applying only to the third generation.  If we tune a modified renormalizable Yukawa coupling $q \tilde{H} b^c$ against the operator $q \tilde{H} b^c H^\dagger H$, we can keep the bottom mass fixed while generating large $b\bar{b}h$ and $b\bar{b}hh$ couplings.  While one could study the interference between these two couplings in Higgs pair production, we know of no well-motivated model to justify the necessary tuning.   For a more motivated scenario, the Yukawa coupling $q \tilde{H} b^c$ could be eliminated altogether, and the bottom mass could come solely from ${\mathcal O}_{FN}$.  This is natural in a two Higgs doublet model, and the smallness of the bottom mass relative to the top would be explained by the fact that symmetries force the bottom coupling to the Higgs to come from a dimension six operator.  However, in this scenario with $\tan \beta = 1$, the $b\bar{b} \rightarrow h h$ cross section at the LHC is on the order of 10 ab, too small to be relevant.

Thus, ${\mathcal O}_{1}$ and ${\mathcal O}_{2}$ are selected out as particularly interesting operators:  these new contributions to TeV-scale physics are not currently constrained by experimental searches, but they have the potential to induce novel physics in Higgs boson pair production at the LHC.  Even if the LHC experiments do not find any new resonances beyond the Higgs boson, high luminosity studies of the Higgs properties at the LHC could still offer a glimpse of the ultraviolet.

\section*{Acknowledgements}
The authors would like to thank Nima Arkani-Hamed for numerous useful
conversations, Hitoshi Murayama for constructive comments, and Johan Alwall, Rikkert Frederix, Fabio Maltoni, and
Simon de Visscher for patient assistance with MadGraph.  A.P. is
supported in part by the MCTP and in part by the DOE.  
J.T. is supported by a fellowship from the Miller Institute for Basic Research 
in Science. L.-T. W. is supported by the DOE under contract DE-FG02-91ER40654. 
L.-T. W. thanks the Aspen Center for Physics for hospitality while this work 
was completed.

\end{document}